\documentclass{emulateapj}
\usepackage{graphicx}

\shorttitle{Strong Gravitational Lens Modeling with Spatially Variant PSFs}
\shortauthors{Rogers \& Fiege}

\begin{document}
\textheight 9.0in

\title{Strong Gravitational Lens Modeling with Spatially Variant Point Spread Functions}
\author{Adam Rogers \& Jason D. Fiege}
\affil{Department of Physics and Astronomy, The University of Manitoba, Winnipeg, Manitoba, R3T-2N2, Canada}
\email{rogers@physics.umanitoba.ca}
\slugcomment{Published in ApJ, 743:68, Dec. 10, 2011. Submitted 2011 July 21; accepted 2011 Sept. 30}

\begin{abstract}
Astronomical instruments generally possess spatially variant point-spread functions, which determine the amount by which an image pixel is blurred as a function of position. Several techniques have been devised to handle this variability in the context of the standard image deconvolution problem. We have developed an iterative gravitational lens modeling code called Mirage that determines the parameters of pixelated source intensity distributions for a given lens model. We are able to include the effects of spatially variant point-spread functions using the iterative procedures in this lensing code. In this paper, we discuss the methods to include spatially variant blurring effects and test the results of the algorithm in the context of gravitational lens modeling problems.
\end{abstract}

\section{Introduction}
\label{sec:intro}

Modeling gravitational lens systems is fundamentally a two step process. A successful model requires an outer optimization procedure to discover the parameters of the lens model, and a nested optimization step to determine the structure of the lensed source object. This inner source optimization step must incorporate the effect of the point spread function (PSF). Observed data are blurred by the PSF, and the presence of noise complicates the deconvolution process in general \citep{hansen}. This problem has been well studied by a variety of authors using spatially invariant PSFs (\citet{WD03}; \citet{K05}; \citet{suyu06}).

To include blurring and lensing effects, we describe operations on images by the application of linear operators. We use ``flattened'' images to facilitate this notation, in which each column of the image is stacked upon the next, forming a vector. Our preferred scheme for solving the lens modeling problem is a modification of the semilinear method of \citet{WD03}, as outlined in \citet{rogersFiege}. To begin, we define image and source coordinate systems that are connected by the thin lens equation:
\begin{equation}
\mbox{\boldmath $\beta$}=\mbox{\boldmath $\theta$}-\mbox{\boldmath $\alpha$} \left(  \mbox{\boldmath $\theta$}  \right),
\label{eq:thinLens}
\end{equation}
where {\boldmath $\theta$} and {\boldmath $\beta$} are the image and source coordinates respectively, and {\boldmath $\alpha$}$(${\boldmath $\theta$}$)$ is the deflection angle determined by the gravitational potential of the lens density distribution. The lens equation is a nonlinear equation that maps pixels from the image to the source plane (\citet{sef}; \citet{wam}).

The source plane pixels are represented as a vector {\boldmath $s$}, with elements that can be varied independently to account for the details of the unknown source intensity distribution. The details of the PSF are encoded in the blurring matrix {\boldmath $B$} and gravitational lens effects, described by Equation \ref{eq:thinLens}, are included in the lensing matrix {\boldmath $L$}. We then define the total lens matrix {\boldmath $f=BL$} and data vector {\boldmath$d$}. Denoting the standard deviation of the noise as $\sigma$, and given a source intensity distribution, the resulting $\chi^2$ statistic between the data and model is
\begin{equation}
\chi^2=\sum_i \frac{\left(d_i-\sum_j f_{ij}s_{j}\right)^2}{\sigma_{i}^2}
\end{equation}
By requiring derivatives of this equation with respect to the elements of the source vector {\boldmath $s$} vanish, we find that the optimal source pixel intensities satisfy a least-squares equation:
\begin{equation}
\mbox{\boldmath $F$}^T \mbox{\boldmath $Fs$}  = \mbox{\boldmath $F$}^T \mbox{\boldmath $\hat{d}$}
\label{eq:LS}
\end{equation}
where we have absorbed factors of $\sigma_i$ into the matrix $F_{ij}=f_{ij}/ \sigma_{i}$ and data vector $\hat{d}_{i}=d_{i}/ \sigma_{i}$. The details of this derivation can be found in \citet{WD03} and \citet{K05}, where the system is solved by direct matrix inversion with regularization. This least squares form is commonly found in the context of large scale image deconvolution problems, which are typically solved by iterative methods (\citet{golub}; \citet{bjorck}; \citet{hansen2}).  Note that the semilinear method provides a technique for solving the linear parameters of the lensed system only (the source intensity distribution) in an ``inner loop'', while the nonlinear parameters of the lens mass distribution must be solved in a separate ``outer loop'' optimization step \citep{rogersFiege}.

Spatial dependence of the PSF is not considered in most conventional deconvolution problems. This simplifies the construction of the blurring matrix {\boldmath $B$}, since only one PSF is taken into account. However, it is well known that the PSF cannot always be treated as constant over an image in cases of astronomical interest. For example, spatially variant PSFs have been studied in the context of adaptive optics (\citet{lauer}; \citet{gilles}) and the PSF of astronomical instruments, such as the Hubble Advanced Camera for Surveys, can be extremely position dependent \citep{bandara}. Several schemes have been designed to deal with this variability (\citet{boden}; \citet{biretta}; \citet{adorf}; \citet{lauer}). Describing a spatially variant PSF is much more complicated than for the invariant case, since each row of the blurring matrix {\boldmath $B$} will be derived from a unique PSF in general. The position of a pixel in the image determines the amount by which it is blurred.

\begin{figure}
\scriptsize \epsscale{1} \plotone{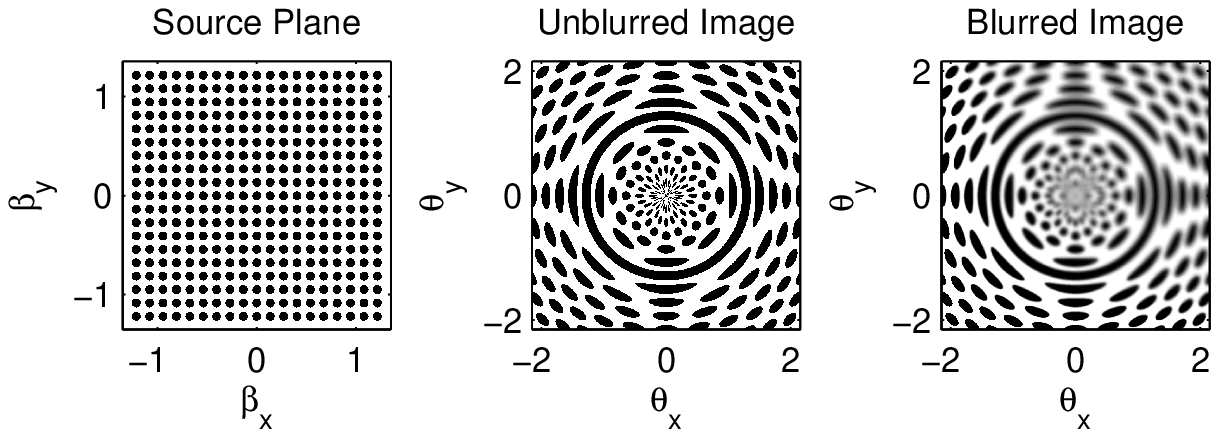}
\caption{Example of spatially variant blurring. Left: a set of regular disks with radius $0.268$  tile the source plane. Center: the circular disks are seen under the lensing effect of a Singular Isothermal Sphere (SIS) lens model. The SIS distorts the background circles into arcs, and the disk at the center of the SIS becomes a complete ring. Right: the same disk pattern under the effect of the SIS lens, with a spatially variant PSF blurring the observation. The blur is described by a delta function in the lower left hand corner to a Gaussian with standard deviation $\sigma_g=6.0$ pixels in the upper right corner, introducing a significant blur.} \label{fig1}
\end{figure}

We illustrate the effect of spatially dependent PSFs on gravitationally lensed images in Figure \ref{fig1}. Consider the lensing effect produced by a singular isothermal sphere (SIS), which has three parameters: velocity dispersion $\sigma_v$ and lens center $\left(x,y\right)$. The deflection angle due to a SIS lens has a simple analytical form most conveniently described in standard polar coordinates $\left( r, \omega \right)$:
\begin{equation}
\alpha\left( r \right)=4\pi\left(\frac{\sigma_v}{c}\right)^2\frac{D_{ls}}{D_{os}},
\label{einsteinRadius}
\end{equation}
where $D_{ls}$ and $D_{os}$ are the angular distances between lens and source and observer and source respectively, and $c$ is the speed of light. We model the blurring in Figure \ref{fig1} with $\sigma_v=265$ km s$^{-1}$ and use source redshift $z_s=1.5$ and lens plane redshift $z_l=0.12$. This model was calculated using cosmological parameters $H_0=70$ km s$^{-1}$ Mpc$^{-1}$, $\Omega_0=0.3$ and $\Lambda_0=0.7$ which we adopt for the remainder of this study. The source is comprised of a set of circular disks in the source plane as shown in the left-hand panel of Figure \ref{fig1} and the gravitationally lensed image is shown in the center panel. The lensed image is then blurred by a spatially variant PSF and is shown in the right panel of Figure \ref{fig1}. The distortion used to create this image varies from a delta function in the lower-left corner (negligible blur) to a Gaussian with standard deviation $\sigma_g=6.0$ pixels in the upper right corner. Each PSF is defined on an arbitrary $33 \times 33$ grid and is normalized to unity sum. The source and image plane size are $240 \times 240$ pixels.


Unlike constant PSFs, spatially variant PSFs cannot be described by a simple convolution operation. Fortunately, numerical methods have been devised to handle them, including sectioning methods \citep{trussel}, which deconvolve each PSF independently and forms the source from the sum of the reconstructions. \citet{nagyOLeary} devised a clever method to model the effects of spatially variant PSFs within the framework of the standard image deconvolution problem. This approach differs from sectioning methods in that the separate PSFs are used to build an approximation to the blurred image of a given source, and a single iterative deconvolution operation is needed to solve for the source intensity distribution. The method was implemented in \citet{nagyRestoreTools} and represents the spatial dependence of the PSF as a summation of piecewise blurring matrices, each of which applies over a limited area of the image. In this study, we use the method of \citet{nagyRestoreTools} to incorporate spatially variant blurring into our gravitational lens modeling code. We briefly review the method here and discuss the procedure in detail in the Appendix.

To include the effects of spatially variant blurs, the image of the unblurred lensed source is padded to enforce a boundary condition \citep{hansen}. We focus on the use of reflexive boundary conditions, in which the image is padded by symmetric reflections of itself. Reflexive boundary conditions tend to reduce ringing artifacts if a significant amount of structure is located near the edges of the image. The image is then divided into a square grid, where the PSF is assumed constant in each region. These image regions and the PSFs are then padded to match in size. The two-dimensional fast Fourier transform (FFT) is used to calculate the resultant blurred image regions independently, resulting in an effective piecewise convolution. By substituting efficient algorithms for the explicit matrix and matrix-transpose multiplications in Equation \ref{eq:LS}, the least squares form of the problem is preserved and the system can be solved efficiently.

In principle spatially variant blurring can be described by a blurring matrix compatible with the semilinear method. However, in practice there are several problems with the matrix approach. First, the size of the blurring matrix is $N_{pix} \times N_{pix}$, so the matrix quickly becomes large as the image resolution is increased. Second, since the PSFs vary over regions of the image, it is possible that {\boldmath $B$} may contain a large number of small but non-zero entries, particularly for large, complicated PSFs that are not well approximated by Gaussians or other simple analytical functions. This complicates the optimization because {\boldmath $M$}$=${\boldmath $F$}$^T${\boldmath $F$} must be inverted in the semilinear scheme. It is generally required that {\boldmath $M$} is sparse in order to store and invert this large matrix. The sparsity requirement helps to reduce computation time and reduces the amplification of noise in the reconstructed source. In practice the semilinear method requires regularization to control the amount of noise present in the solution of Equation \ref{eq:LS}. The details and effects of several distinct regularization methods used with the semilinear method were studied in detail by \citet{suyu06}.

Our previous work \citep{rogersFiege}, compared the semilinear method with several iterative methods to solve the least-squares problem (Equation \ref{eq:LS}). Iterative schemes have the advantage that time is saved by avoiding the explicit construction of the lens and blurring matrices. This is done using direct interpolation on the source plane under the effect of the lens equation (Equation \ref{eq:thinLens}).

\citet{rogersFiege} used the Qubist Optimization Toolbox \citep{qubist} to map the $\chi^2$ surface over the space of the nonlinear lens parameters using the Ferret Genetic Algorithm (GA) and Locust Particle Swarm Optimizer (PSO). Since this mapping requires a large number of function evaluations ($\approx 10^5$) over the course of a run, speed is of the essence when choosing an inner loop optimization to determine the source plane parameters.

Using the techniques introduced by \citet{nagyRestoreTools} as a foundation, we have added the capability to include spatially variant PSFs to our gravitational lens modeling code using piecewise constant PSFs. This new capability is the subject of the current exploration.


\section{A Small-Scale Test}
In this section, we provide an example of modeling an extended source under the effects of a spatially variant PSF. We generate the lensed image of an analytical function that describes a spiral source intensity distribution, according to the equation:
\begin{equation}
S(r,\omega)=\frac{S_0}{r_c^2+r^2} \exp\left[-2 \sin^2\left(\omega-\omega_0-\tau r^2\right)\right],
\label{bonnetSource}
\end{equation}
where $S_0$ is the maximum brightness in arbitrary units and core radius $r_c$.  The tightness of the arms about the central bulge is controlled by $\tau$, and $\omega_0$ controls the orientation of the spiral, in standard polar coordinates $\left(r,\omega\right)$. This artificial ``galaxy'', originally described by \citet{bonnet}, serves as a convenient test pattern. To draw comparisons between our results for spatially invariant PSFs \citep{rogersFiege}, we will again make use of a Singular Isothermal Ellipse (SIE; \citet{keeton}) with deflection angle components:

\begin{equation}
\alpha_x=\frac{bq}{\sqrt{1-q^2}}\tan^{-1}\left( \frac{x\sqrt{1-q^2}}{\psi+s} \right)
\label{SIEX}
\end{equation}

\begin{equation}
\alpha_y=\frac{bq}{\sqrt{1-q^2}}\tanh^{-1}\left( \frac{y\sqrt{1-q^2}}{\psi+q^2s} \right),
\label{SIEY}
\end{equation}
with $\psi^2=q^2(s^2+x^2)+y^2$ and $q=\sqrt{ (1-\epsilon)/(1+\epsilon)}$, and $b$ is the equivalent Einstein radius when $q=1$.  The parameter $b$ is related to the velocity dispersion $\sigma_v$ by Equation \ref{einsteinRadius}.

The parameters used in this test are velocity dispersion $\sigma_v=265$ km s$^{-1}$ with $z_d = 0.3$ and $z_s = 1.05$ giving an equivalent Einstein ring of $b=1.32$ arcsec, ellipticity $\epsilon=0.35$, lens center $(x,y)=(0.11,0)$, core size $s$, and orientation angle $\theta_L=\pi/4$ measured counterclockwise from the right of the image.  We set $s=0$, resulting in a singular mass distribution.

\begin{figure}
\scriptsize \epsscale{1} \plotone{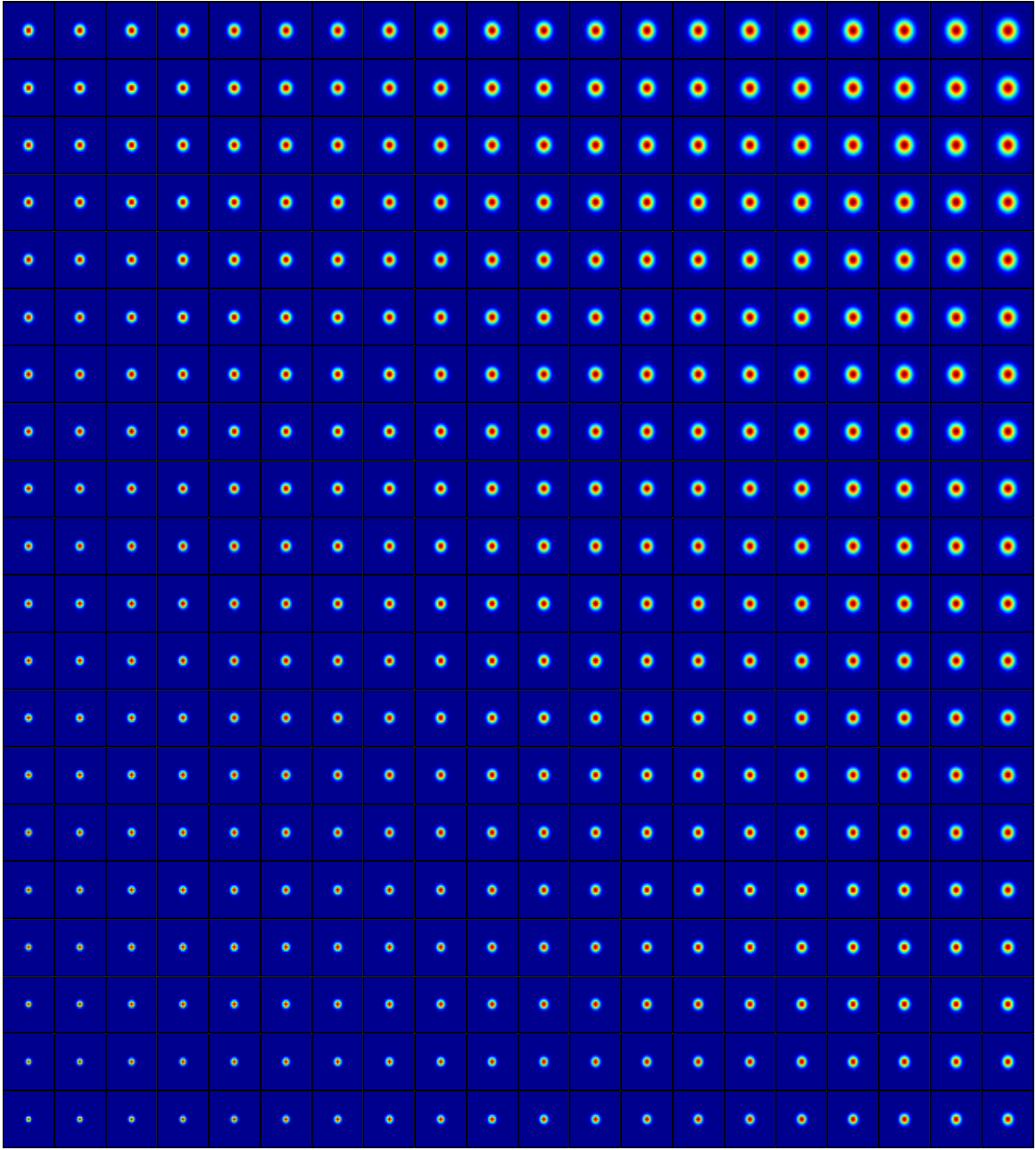}
\caption{Grid of PSFs used in Figure \ref{fig2image}. The PSFs vary from a Gaussian of FWHM of $2.35$ pixels in the lower-left corner producing a modest blur to a Gaussian with FWHM $4.75$ pixels in the upper right corner.}
\label{psfGrid}
\end{figure}

\begin{figure}
\scriptsize \epsscale{1.10} \plotone{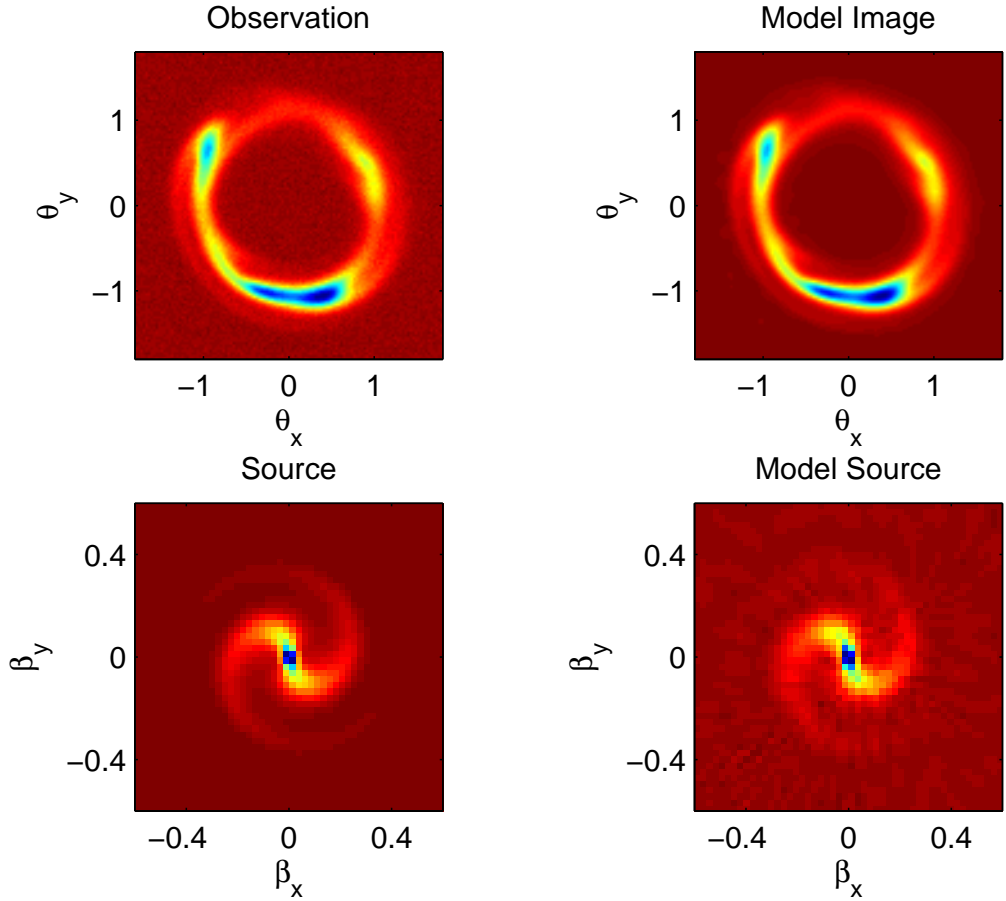}
\caption{Top left: artificial data on a $120 \times 120$ grid. Bottom left: artificial source on a $50 \times 50$ grid. Top right: model observation. Bottom right: model source. The results after $19$ iterations are shown. Note the presence of reconstructed noise in the source. The model has a reduced $\chi^2=0.998$.}
\label{fig2image}
\end{figure}

\begin{figure}
\scriptsize \epsscale{0.75} \plotone{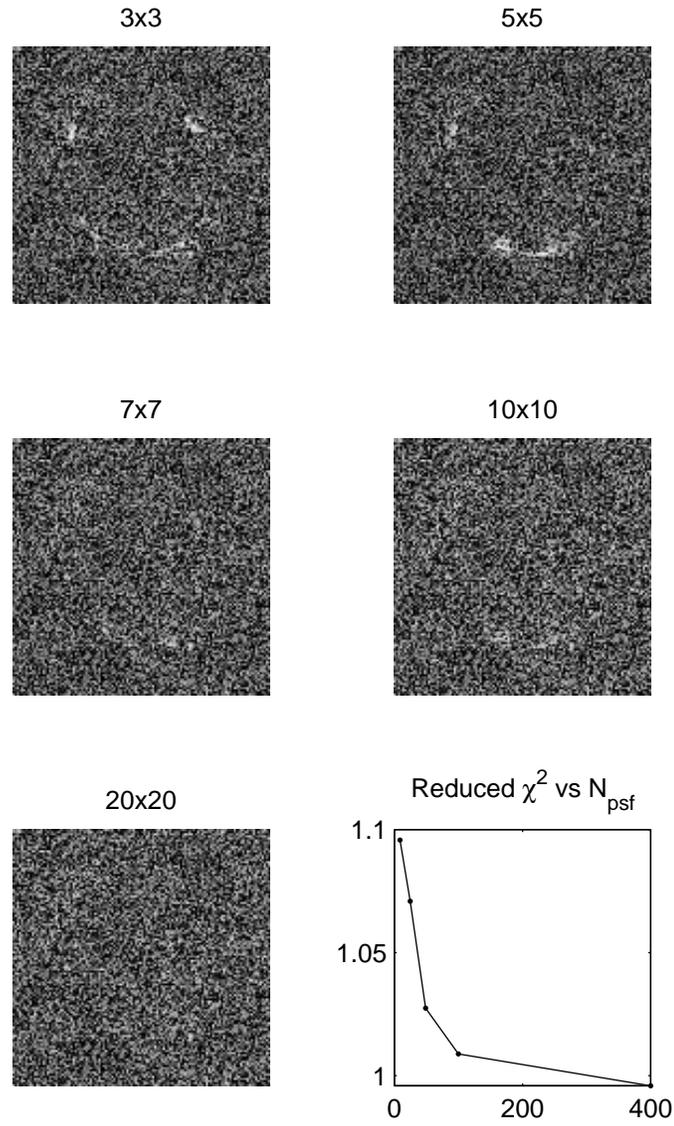}
\caption{Image residuals for a $3\times 3$, $5\times 5$, $7\times 7$, $10\times 10$, and $20\times 20$ PSF grids after $19$ CGLS iterations. For a small number of PSFs there is a significant amount of residual structure, but these artifacts are reduced as the grid of PSFs is enlarged. The reduced $\chi^2$ is shown as a function of the number of PSFs ($N_{psf}$) used in the inversion.}
\label{f2residual}
\end{figure}

We used Equation \ref{eq:thinLens} to form the lensed image of the source (Equation \ref{bonnetSource}) using the SIE deflection angle formulae. We generated a $20 \times 20$ grid of spatially variant Gaussian PSFs where each PSF is defined by a $33\times 33$ pixel mesh and has an FWHM ranging from $2.35$ to $4.8$ pixels, shown in Figure \ref{psfGrid}.
This grid of PSFs was used to blur the gravitationally lensed image and additive Gaussian white noise with standard deviation $\sigma_g=1.05$ was added after the blurring operation, resulting in the artificial data shown in Figure \ref{fig2image}. We define the peak signal-to-noise ratio (PSNR) as
\begin{equation}
PSNR=\frac{I_{max}}{\sigma_g},
\end{equation}
giving $PSNR=105.83$. To illustrate the effect of varying the number of PSFs, we model the data using smaller grids of $3\times 3$, $5\times 5$, $7\times 7$, $10\times 10$, and $20\times 20$ PSFs. As shown in Figure \ref{f2residual}, the best reconstruction with the lowest reduced $\chi^2$ is obtained using a grid of $20\times 20$ PSFs, which is the same number used to generate the data. This source and corresponding model image after $20$ iterations are also shown in Figure \ref{fig2image}. The $3\times 3$ and $5\times 5$ image residuals show significant structure, which is not present in the finer approximations. The residuals using a grid of $20 \times 20$ PSFs appear featureless. This demonstrates the improvement in image reconstruction as we include successively more information characterizing the blur.

\begin{figure}
\scriptsize \epsscale{1.00} \plotone{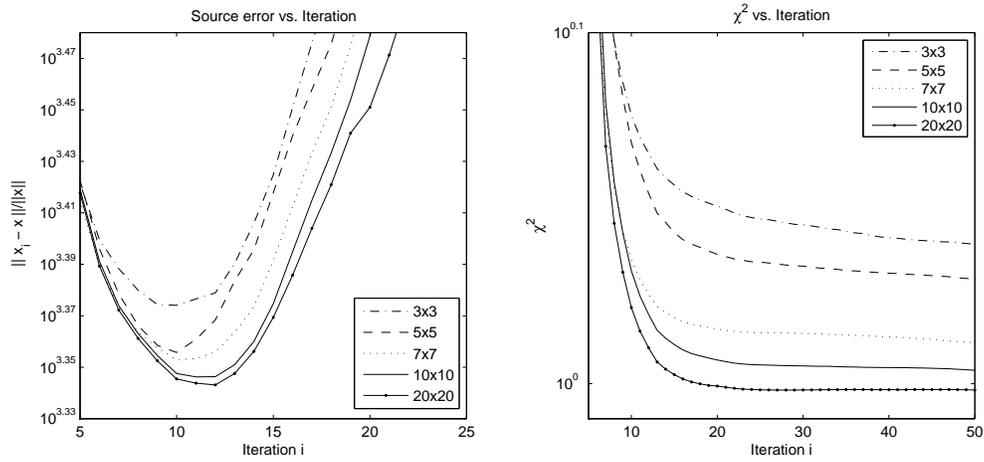}
\caption{Left: source convergence history using the CGLS algorithm. Right: corresponding Image convergence history. Note that the source displays semi-convergent behavior. The disagreement between model and actual source reaches a minimum before increasing. The critical iteration changes as the PSF grid is enlarged.}
\label{fig2converge}
\end{figure}

\begin{figure}
\scriptsize \epsscale{1.00} \plotone{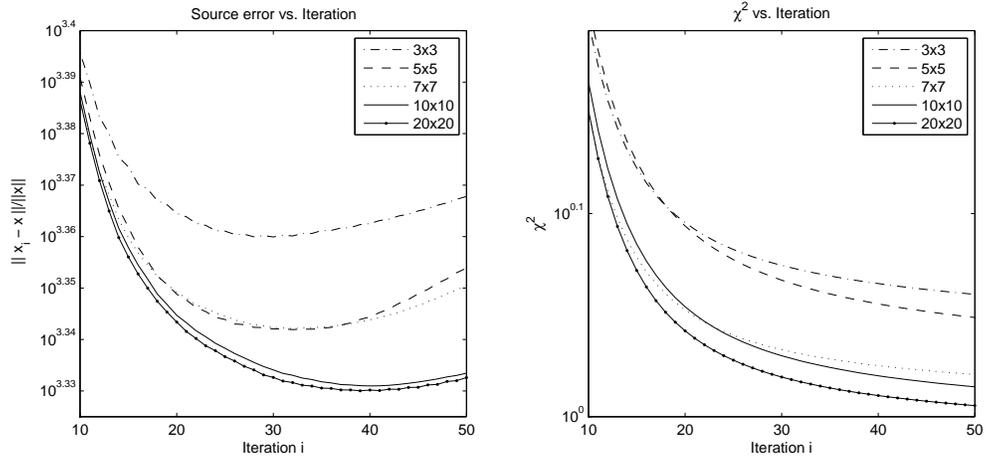}
\caption{Left: source convergence history using the SD algorithm. Right: corresponding Image convergence history. The semi-convergent behavior of the source is less extreme than for the CGLS algorithm.}
\label{fig2SD}
\end{figure}

Figure \ref{fig2converge} shows the relative error between the model source and the true solution as a function of iteration. For all PSF grid sizes, we find that the solutions display semi-convergence behavior such that the relative error between the model solution and the true solution improves until a minimum is reached and then begins to increase. This is due to the properties of the local optimizer used to determine the optimal source, and arises in the deconvolution step due to noise in the observed image. Regularization methods are generally used to control the increase of noise in the reconstructed source found by the semilinear method \citep{suyu06}. Several optimization methods have been applied to problems with spatially variant blur including Landweber iteration (\citet{nocedal}; \citet{fish};  \citet{trusselHunt}), Richardson-Lucy deconvolution \citep{faisal}, and Lanczos-Tikhonov hybrid methods \citep{hybr} in the context of the standard image deconvolution problem. Following \citet{rogersFiege}, we focus on the conjugate gradient method for least-squares problems (CGLS) and the steepest descent method (SD). Figure \ref{fig2SD} shows the convergence history of the SD algorithm. As in the invariant PSF case discussed in \citet{rogersFiege}, the SD solution converges more slowly than CGLS and therefore it is less sensitive to the stopping criteria. When using an iterative method for local optimization, the number of iterations itself acts as a regularization parameter. The optimal stopping iteration of these local optimizers is at the minimum of the relative error curve for a given set of lens parameters and PSF tiling. This critical iteration represents a balance between the reduced image $\chi^2$ and the amount of regularization used \citep{press2007}. Established methods exist to determine this critical iteration, including the L-Curve criterion \citep{hansenOLeary} and Generalized Cross Validation \citep{golubGCV}. In previous work \citep{rogersFiege} we made use of the L-Curve criterion but Generalized Cross Validation is also implemented in our software.

\begin{figure}
\scriptsize \epsscale{1} \plotone{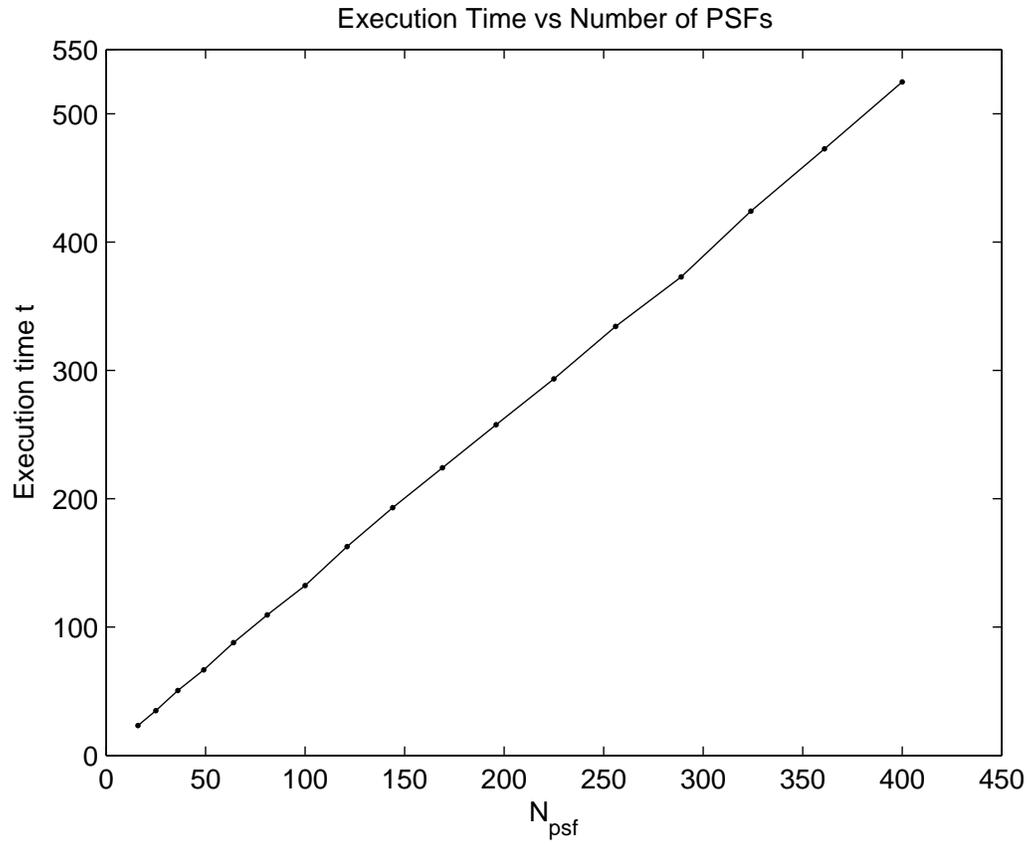}
\caption{Timing results for the CGLS algorithm using $N_{psf}$ as the number of PSFs to approximate the blurring effect. The plot illustrates the runtime for $4 \times 4$ to $20 \times 20$ square PSF grids in seconds. Each CGLS run was terminated at $20$ iterations.}
\label{timing}
\end{figure}

We find that the execution time of the problem including a spatially variant PSF increases approximately linearly with the number of separate PSFs used in the inversion as shown in Figure \ref{timing}. This suggests that significant gains could be made in the efficiency of the routine by parallelizing the implementation, since each image region is independent. By splitting up the problem over several processors, the runtime for very large PSF grids can become feasible.

\section{A Large-Scale Test}

To demonstrate the code in operation on a large scale problem, we simulate the lensing effect of the mass distribution of a galaxy cluster on a portion of the Hubble deep field using an elliptical potential. This test is intended as a demonstration of the feasibility and efficiency of our method on a problem that would be difficult using the semilinear method while including a spatially dependent PSF. Problems of this size are realistic for a number of practical modeling situations. For example, \citet{alard} has modeled the lensed system SL2SJ021408-053532, which produces a set of large arcs. This system has a lens that is comprised of a small group of six galaxies. Due to the large size of the lensed arcs, the scope of the source modeling prohibited the direct application of the semilinear method.

We form the lensed image of a portion of the Hubble deep field \citep{williams} by applying the elliptical potential of \citet{blandford} which was used by \citet{link} to simulate the lens effect of the dark matter distribution of galaxy clusters. This potential function is given by
\begin{equation}
\psi(x,y)=\frac{b^{2(1-q)}}{2q}\left[ s^2+(1+\epsilon_c)x^2+2\epsilon_sxy+(1-\epsilon_c)y^2 \right]^q,
\label{psiPot}
\end{equation}
which results in deflection angle {\boldmath $\alpha$}$(${\boldmath $\theta$}$)=\nabla \psi(${\boldmath $\theta$}$)$. The elliptical potential depends on seven parameters: $b$ is the equivalent Einstein radius in the limit of vanishing core radius $s$, ellipticity $\epsilon$, and power law index $q$, where $0 \leq q \leq 0.5$. The position angle of the lens $\phi$ determines the functions $\epsilon_c=\epsilon \cos\phi$ and $\epsilon_s=\epsilon \sin\phi$. We use the Einstein radius $b=9$, power law index $q=0.25$, $\phi=\pi/4$, position $(x,y)=(0,0)$ and $s=0.5$. The lens and source redshifts are $z_d=0.12$ and $z_s=1.5$, respectively. We used an array of $25$ PSFs arranged on a $5\times 5$ grid to blur the image. This set of PSFs has been used to test image restoration schemes for Hubble Space Telescope (HST) images and represents the spatially variant nature of the aberrations affecting the HST before it was repaired (\citet{kat}; \citet{nagyOLeary}). The size of each PSF is $60\times 60$ pixels, and the source and image plane used to generate our lensed image are $800 \times 800$ pixel$^2$. Gaussian white noise was added with standard deviation $\sigma_g=1.37$, giving the image $PSNR=138.4$.

\begin{figure}
\scriptsize \epsscale{1} \plotone{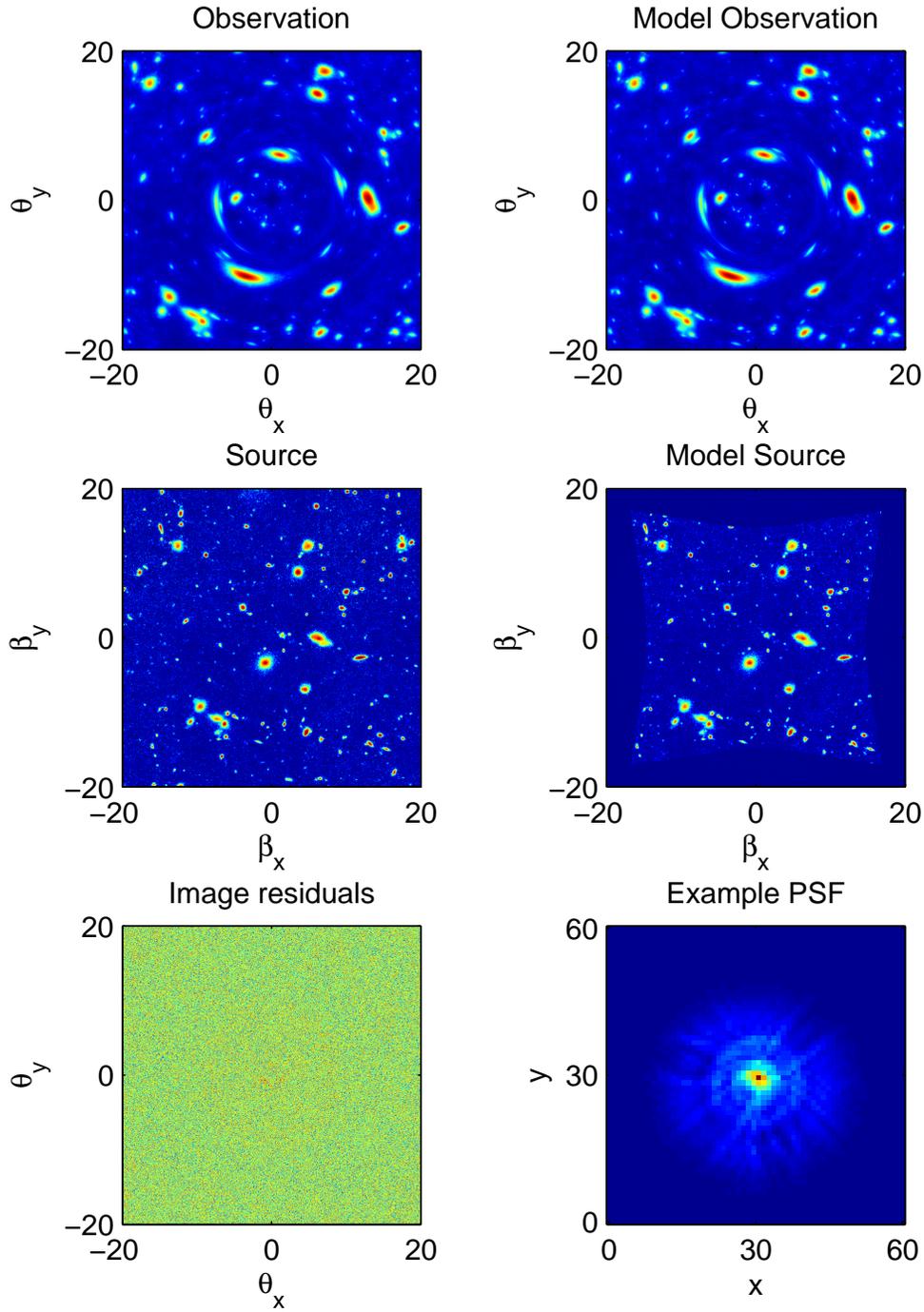}
\caption{Top row: observation and model image. Middle row: actual and model source. The image and source plane are both $800 \times 800$ pixels. These results are shown for $100$ iterations. Bottom row: image residuals and an example of one of the $25$ large PSFs used to generate the observations. Both of these images are plotted in logarithmic intensity to emphasize low level structure. Approximate runtime for this large-scale test is approximately $7$ minutes.}
\label{fig3cluster}
\end{figure}

The image after $100$ iterations is shown in Figure \ref{fig3cluster}, and a reduced $\chi^2=0.995$ was found. The system was solved using the CGLS algorithm with all $25$ PSFs using the lens parameters defined above. The model took approximately $7$ minutes to solve using a single 2.4 GHz CPU core. An approximation to the nonlinear lens parameters could be found using global optimization methods if one of the following strategies were employed: (1) a low-resolution approximation to the data could be used early during the lens parameter optimization, with successive refinement occurring later during the run; (2) a global optimizer could be used to roughly approximate the lens parameters, shifting to a faster local optimization scheme once solutions are localized to a small region of parameter space; or (3) global optimization could be used for the entire problem making use of large-scale parallelization.

\section{Conclusion}
We have developed a method to include the effects of a spatially variant PSF in gravitational lens modeling. Including these effects in the standard semilinear method would be difficult due to the complicated blurring matrix required. These complications can be overcome easily by incorporating the method of \citet{nagyRestoreTools}. Our approach can accommodate large lensing problems like the case studied by \citet{alard}, which limits the applicability of the direct semilinear approach. Techniques to include the effects of spatially variant PSFs are important, as the response varies over the detector area for many astronomical instruments. Our algorithm allows this effect to be included in lensing problems, thus improving the quality of reconstructions when the variability of the PSF is significant. The CGLS and SD algorithms allow a regularized inversion to be found quickly by truncated iteration.

\section{Acknowledgements}
A.R. acknowledges NSERC for funding this research, and J.F. acknowledges funding from an NSERC Discovery Grant.  The authors also thank the anonymous referee, whose comments helped to improve the flow of the paper. 

\appendix
\section{Spatially Variant Blurring Effects}

To describe blurring by a spatially variant PSF we first present an efficient method using two-dimensional FFTs. We then show how to treat the problem in terms of blurring matrices and flattened image vectors. See \citet{nagyOLeary} for more details on the approach and \citet{nagyRestoreTools} for a MATLAB implementation.

Consider an $N \times N$ grid of independent PSFs {\boldmath $P$}$_{ij}$ and split the unknown blurred image {\boldmath $Y$} into regions {\boldmath $Y$}$_{ij}$, each of size $k \times k$:
\begin{equation}
\mbox{\boldmath $Y$}=
\begin{tabular}{|l|l|l|l|}
\hline
{\boldmath $Y$}$_{11}$ & {\boldmath $Y$}$_{12}$ & $\cdots$ & {\boldmath $Y$}$_{1N}$ \\ \hline
{\boldmath $Y$}$_{21}$ & {\boldmath $Y$}$_{22}$ & $\cdots$ & {\boldmath $Y$}$_{2N}$ \\ \hline
$\vdots$  & $\vdots$  & $\ddots$ & $\vdots$  \\ \hline
{\boldmath $Y$}$_{N1}$ & {\boldmath $Y$}$_{N2}$ & $\cdots$ & {\boldmath $Y$}$_{NN}$\\ \hline
\end{tabular}
\end{equation}
Each of these blocks will be affected by an independent PSF. Suppose that the size of each PSF is $(r+1) \times (r+1)$ with $r$ even, and let the unblurred $N \times N$ image be represented by {\boldmath $X$}. 

Let us define a set of ``mask'' matrices {\boldmath $w$}$_{ij}$. In the case of piecewise constant PSFs, these masks are the same size as the unblurred image and are comprised of $0$ entries everywhere except for the $k \times k$ block at position $(i,j)$, where the entries of {\boldmath $w$}$_{ij}$ are set to $1$. 

To find the components of a given region we convolve {\boldmath $X$} with the corresponding PSF $P_{ij}$, followed by an element-wise multiplication by the mask {\boldmath $w$}$_{ij}$. The non-zero elements of this product give {\boldmath $Y$}$_{ij}$. Proceeding in this way we build up the blurred image block by block:
\begin{equation}
\mbox{\boldmath $Y$}_{ij} = \sum_{i=1}^{N}\sum_{j=1}^{N} \mbox{\boldmath $w$}_{ij} \circ \left( \mbox{\boldmath $P$}_{ij} \ast \mbox{\boldmath $X$} \right),
\label{eq:convOS}
\end{equation}
where the symbol ``$\circ$'' represents element-wise multiplication and symbol ``$\ast$'' is the convolution operation. Note that each term in the sum is determined by the convolution of the entire image {\boldmath $X$} with the appropriate PSF before the mask is applied. This is crucial to ensure that ``seams'' will not be visible between regions in the blurred image {\boldmath $Y$}.

In general, it is possible to speed up this routine by calculating {\boldmath $Y$}$_{ij}$ directly. Consider splitting the unblurred image into regions {\boldmath $X$}$_{ij}^{k}$ where the superscript denotes the size of the block, in this case $k \times k$. In order to avoid artifacts and keep the correct intensity near the edges of this block after convolution, we include a number of neighboring rows and columns on each side of {\boldmath $X$}$_{ij}^{k}$. The width of this border is set by the size of the PSF, $r/2$, with regions on the image boundary padded to enforce the boundary conditions discussed in Section \ref{sec:intro}. These extended regions are then denoted {\boldmath $X$}$_{ij}^{(r+k)}$. The PSFs are padded to match the extended regions in size, resulting in {\boldmath $P$}$_{ij}^{(r+k)}$. The blurred extended region is found by the convolution
\begin{equation}
\mbox{\boldmath $Y$}_{ij}^{(r+k)} = \left( \mbox{\boldmath $P$}_{ij}^{(r+k)} \ast \mbox{\boldmath $X$}_{ij}^{(r+k)} \right).
\end{equation}
The central $k \times k$ block of this product is clipped out and placed in the $(i,j)$ position of {\boldmath $Y$}. The process is repeated until the entire blurred image is filled in. Time is saved working with extended regions and padded PSFs since we only need to calculate the convolution over the $(r+k) \times (r+k)$ block for each PSF rather than the entire image as in Equation \ref{eq:convOS}, and the construction of masks is not needed. The convolutions can be carried out efficiently with two-dimensional FFTs. 

The basic procedure can also be described by an analogous matrix-vector operation. To express the sum in Equation \ref{eq:convOS} in terms of matrix multiplication, we define the unblurred flattened image as a vector {\boldmath $x$}, and the flattened blurred image as {\boldmath $y$}. We build a set of $N^2$ blurring matrices to describe the effect of each PSF on {\boldmath $x$}, which we denote as {\boldmath $B$}$_{ij}$. The mask matrices {\boldmath $w$}$_{ij}$ are used to construct analogous weighting matrices {\boldmath $D$}$_{ij}$. These matrices are of size $N_{pix} \times N_{pix}$, where $N_{pix}$ is the number of pixels in the image, identical to the size of the blurring matrices {\boldmath $B$}$_{ij}$. The total blurring matrix {\boldmath $B$} is then written as a weighted sum of blurring matrices {\boldmath $B$}$_{11}$, {\boldmath $B$}$_{12}$,...,{\boldmath $B$}$_{NN}$.
\begin{equation}
\mbox{\boldmath $B$}=\sum_{i=1}^{N} \sum_{j=1}^{N}\mbox{\boldmath $D$}_{ij} \mbox{\boldmath $B$}_{ij}.
\label{eq:svblur}
\end{equation}
The blurred image is then found by a matrix multiplication {\boldmath $y$}={\boldmath $Bx$}. The weighting matrices {\boldmath $D$}$_{ij}$ have the $m$th diagonal entry equal to $1$ provided that image pixel $m$ is in region $(i,j)$, and all other elements $0$. The weighting matrices satisfy $\sum_{i=1}^N \sum_{j=1}^N$ {\boldmath $D$}$_{ij} = ${\boldmath $I$} where {\boldmath $I$} is the $N_{pix} \times N_{pix}$ identity. We adopt the use of piecewise constant PSFs but in general it is possible to include higher order interpolation schemes between PSFs using the weighting matrices. The case of linear interpolation in solving systems with spatially variant blur has been studied by \citet{nagyOLeary}, but its inclusion complicates the procedure and did not provide a significant improvement to the quality of the solution and increased computation times \citep{nagyRestoreTools}.


\clearpage
\newpage

\clearpage
\newpage

\clearpage
\newpage

\clearpage
\newpage

\clearpage
\newpage

\clearpage
\newpage

\clearpage
\newpage

\clearpage
\newpage


\begin{thebibliography}{}
 \bibliographystyle{abbrv}
 \bibliographystyle{references}

\bibitem[Adorf(1994)]{adorf} Adorf, H.M., 1994, in The Restoration of HST Images and Spectra II, ed. R. J. Hanisch \& R. L. White (Baltimore, MD: Space Telescope Science Institute), 72

\bibitem[Alard(2009)]{alard} Alard, C. 2009, A\& A, 506, 609

\bibitem[Bandara et al.(2009)]{bandara} Bandara, K., Crampton, D., \& Simard, L. 2009, ApJ, 704, 1135

\bibitem[Biretta(1994)]{biretta} Biretta, J. 1994 in The Restoration of HST Images and Spectra II, ed. R. J. Hanisch \& R. L. White (Baltimore, MD: Space Telescope Science Institute), 72

\bibitem[Bj{\"o}rck(1996)]{bjorck} Bj{\"o}rck, {\AA}. 1996, Numerical Methods for Least Squares Problems, (Philadelphia, PA: SIAM Publishers)

\bibitem[Blandford \& Kochanek(1987)]{blandford} Blandford, R. D., \& Kochanek, C. S., 1987, ApJ, 321, 658

\bibitem[Boden et al.(1995)]{boden} Boden, A.F., Redding, D.C., Hanisch, R.J., \& Mo, J., 1995, J. Opt. Soc. Am. A, 13, 1537

\bibitem[Bonnet(1995)]{bonnet} Bonnet, H. 1995, PhD thesis, L'Univ. Paul Sabatier de Toulouse


\bibitem[Chung et al.(2008)]{hybr} Chung, J., Nagy, J. G., \& O'Leary, D. P. 2008, Elec. Trans. Numer. Anal., 28, 149


\bibitem[Faisal et al.(1995)]{faisal}Faisal, M., Lanterman, A.D., Snyder, D.L., \& White, R.L. 1995, J. Opt. Soc. Am. A, 12, 2593

\bibitem[Fiege(2010)]{qubist} Fiege, J. D., 2010, Qubist Users Guide: Optimization, Data Modeling, and Visualization with the Qubist Optimization Toolbox for MATLAB (Winnipeg: nQube Technical Computing)

\bibitem[Fish et al.(1996)]{fish} Fish, D. A., Grochmalicki, J., \& Pike, E. R. 1996, J. Opt. Soc. Am. A, 13, 1

\bibitem[Gilles et al.(2002)]{gilles}Gilles, L.,  Vogel, C. R. \& Bardsley, J. M. 2002, Inverse Probl. 18, 237

\bibitem[Golub et al.(1979)]{golubGCV} Golub, G.H., Heath, M., \& Wahba, G. 1979, Technometrics, 21, 2, 215-223

\bibitem[Golub \& Reinsch(1970)]{golub} Golub, G.H., \& Reinsch, C. 1970, Numer. Math., 14, 403

\bibitem[Hansen(2010)]{hansen2} Hansen, P.C. 2010, Discrete Inverse Problems: Insight and Algorithms (Philadelphia, PA: SIAM publishers)

\bibitem[Hansen et al.(2006)]{hansen} Hansen, P.C., Nagy, J.G., \& O'Leary,D.P. 2006, Deblurring Images: Matrices, Spectra and Filtering (Philadelphia, PA: SIAM publishers)

\bibitem[Hansen \& O'Leary(1993)]{hansenOLeary} Hansen, P.C., \& O'Leary, D.P. 1993, SIAM J. Sci. Comput., 14, 1487

\bibitem[Katsaggelos et al.(1994)]{kat} Katsaggelos, A. K., Kang, M. G., \& Banham, M. R. 1994, in The Restoration of HST Images and Spectra II, ed. R. J. Hanisch \& R. L. White (Baltimore, MD: Space Telescope Science Institute), 3.

\bibitem[Keeton \& Kochanek(1998)]{keeton} Keeton, C.R. \& Kochanek, C.S. 1998, ApJ, 495, 157

\bibitem[Koopmans(2005)]{K05} Koopmans, L. V. E. 2005, MNRAS, 363, 1136


\bibitem[Lauer(2002)]{lauer} Lauer, T. 2002, Proc. SPIE, 4847, 167

\bibitem[Link \& Pierce(1998)]{link} Link, R., \& Pierce, M. J., 1987, ApJ, 502, 63

\bibitem[Nagy \& O'Leary(1998)]{nagyOLeary} Nagy, J. G. \& O'Leary, D. P. 1998, SIAM J. Sci. Comput. 19, 1063

\bibitem[Nagy et al.(2002)]{nagyRestoreTools} Nagy, J.G., Palmer, K.M., \& Perrone, L. 2002, Numer. Algorithms, 36, 73


\bibitem[Nocedal \& Wright(1999)]{nocedal} Nocedal, J. \& Wright, S. J. 1999, Numerical Optimization (New York: Springer)

\bibitem[Petters et al.(2001)]{wam} Petters, A.O., Levine, H. \& Wambsganns, J. 2001, Singularity Theory and Gravitational Lensing (Boston, MA: Birkh\"{a}user)

\bibitem[Press et al.(2007)]{press2007} Press, W.H., Teukolsky, S.A., Vetterling, W.T., \& Flannery, B.P. 2007, Numerical Recipes: The Art of Scientific Computing (3rd ed.; New York:Cambridge Univ. Press)

\bibitem[Rogers \& Fiege(2011)]{rogersFiege} Rogers, A. \& Fiege, J. D. 2011, ApJ, 727, 80

\bibitem[Schneider et al.(1992)]{sef}Schneider, P., Ehlers, J., \& Falco, E. E. 1992, Gravitational Lenses (Berlin: Springer)

\bibitem[Suyu et al.(2006)]{suyu06} Suyu, S. H., Marshall, P. J., Hobson, M. P., \& Blandford, R. D. 2006, MNRAS, 371, 983

\bibitem[Trussel \& Fogel(1992)]{trussel} Trussell, H. J. \& Fogel, S. 1992, IEEE Trans. Image Proc. 1, 123

\bibitem[Trussel \& Hunt(1978)]{trusselHunt} Trussell, H. J. \& Hunt, B. R. 1978, IEEE Trans. Acoust. Speech, Signal Processing, 26, 608

\bibitem[Warren \& Dye(2003)]{WD03} Warren, S. J. \& Dye, S. 2003, ApJ, 590, 673

\bibitem[Williams et al.(1996)]{williams} Williams, R. E., Blacker, B., Dickinson, M., et al. 1996, AJ, 112, 1335


\end{thebibliography}
\end{document}